# Excited States of Proton-bound DNA/RNA Base Homo-dimers: Pyrimidines

Géraldine Féraud[1], Matias Berdakin[2], Claude Dedonder[1], Christophe Jouvet[1]* and Gustavo A. Pino[2]*

[1] *CNRS, Aix-Marseille Université, Physique des Interactions Ioniques et Moléculaires (PIIM), UMR 7345 - 13397 Marseille Cedex 20, France*

[2] *INFIQC (CONICET – Universidad Nacional de Córdoba) Dpto. de Fisicoquímica – Facultad de Ciencias Químicas – Centro Láser de Ciencias Moleculares – Universidad Nacional de Córdoba, Ciudad Universitaria, X5000HUA Córdoba, Argentina*

*Corresponding authors:
christophe.jouvet@univ-amu.fr, *CNRS, Aix-Marseille Université, Physique des Interactions Ioniques et Moléculaires (PIIM), UMR 7345 - 13397 Marseille Cedex 20, France,* +33413594610

gpino@fcq.unc.edu.ar, *INFIQC (CONICET – Universidad Nacional de Córdoba) Dpto. de Fisicoquímica – Facultad de Ciencias Químicas – Centro Láser de Ciencias Moleculares – Universidad Nacional de Córdoba, Ciudad Universitaria, X5000HUA Córdoba, Argentina,* +54-351-5353565

## Abstract

We are presenting the electronic photo fragment spectra of the protonated pyrimidine DNA bases homo-dimers. Only the thymine dimer exhibits a well structured vibrational progression, while protonated monomer shows broad vibrational bands. This shows that proton bonding can block some non radiative processes present in the monomer.

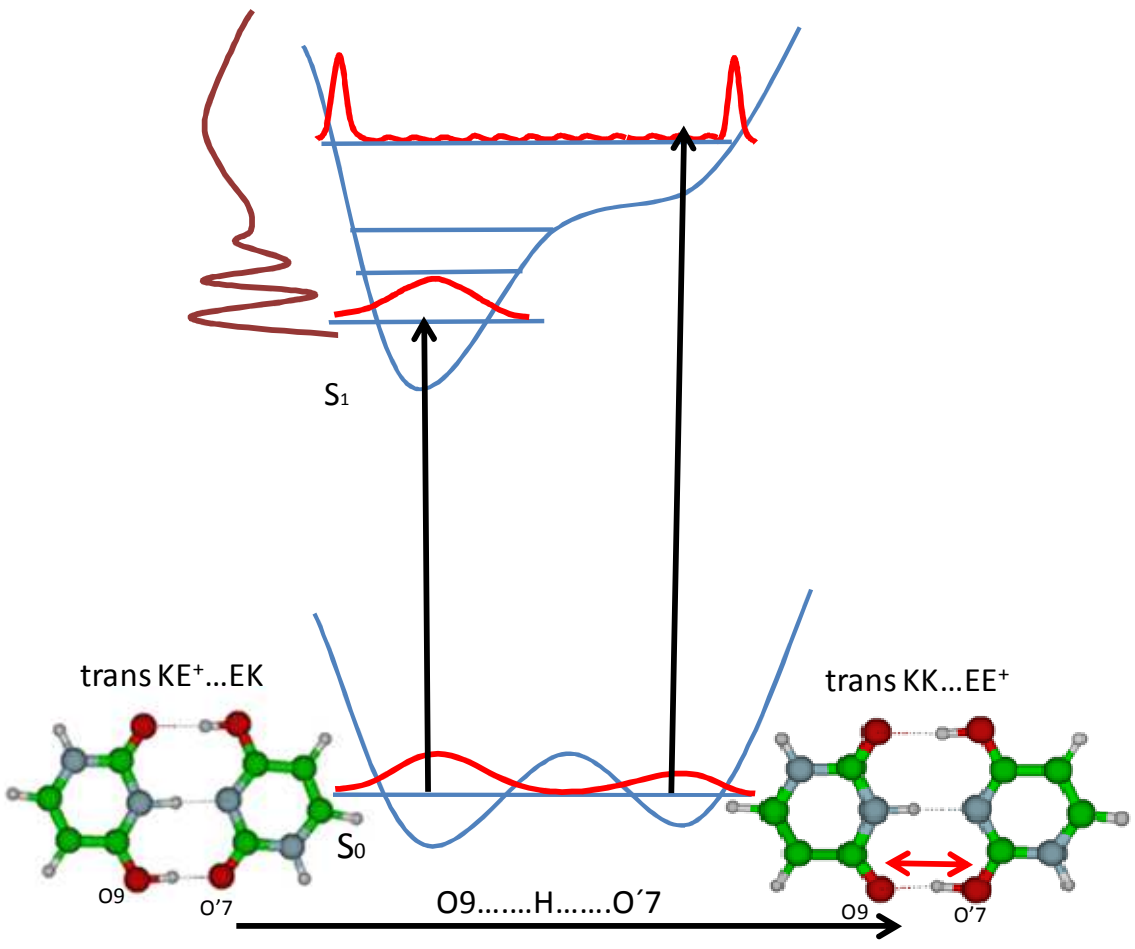

**Keywords:** Proton Transfer – Excited State – Cold ions – Tautomers

## 1. Introduction

Since a few years the effect of the proton on the excited states properties of aromatic molecules as well as clusters has been investigated.[1] Some general trends begin to appear.

(i) For many protonated systems in which the proton is quite far from the aromatic ring (as in aromatic amines and amino acids), the effect of the proton is more important on the excited state lifetime than on the excited state energy of the system (e.g. the excited state lifetime of tryptophan is shortened from the ns to the fs regime[2–4] upon protonation, while the excitation energy only changes by 0.1 eV). The short excited state lifetime in protonated tryptophan has been quite nicely explained by the excited state dynamics of the proton due to the presence of a dissociative $\pi\sigma^*$ leading to proton transfer or H loss.[5]

(ii) For other aromatic molecules such as PAHs (polycyclic aromatic hydrocarbons), indole, benzaldehyde, in which the proton is added to one atom of the aromatic ring, the effect of the proton is particularly important on the excited state energy, inducing a strong red-shift in the electronic absorption as compared to the neutral molecules.[6–9] This was explained by the presence of a charge transfer excited state (CT), in which an electron from the “neutral” moiety of the molecules is promoted to the protonated part of the system.[10] The simplest example is the protonated benzene dimer for which the first electronic transition lies in the visible and corresponds to the excitation of an electron from the HOMO on the neutral benzene toward the LUMO on the protonated benzene moiety.[11]

(iii) We have recently investigated the protonated DNA/RNA bases and found that the electronic transitions are not much shifted as compared to the transitions of their neutral homologues and that the excited state lifetimes are strongly dependant on the tautomeric structure.[12] For example in protonated uracil, the electronic spectrum of the Enol-Keto† ($EK^+$) tautomer exhibits very sharp vibrational structures while the bands are considerably lifetime broadened in the Enol-Enol ($EE^+$) tautomer. Thus, some non-radiative processes are involved in the excited state dynamics of the protonated bases. As there are heteroatoms such as oxygen and nitrogen in DNA/RNA bases, relaxation mechanisms could include the proton/hydrogen loss or hindered loss, mediated by a $\pi\sigma^*$ dissociative state as in protonated aromatic amines (including amino acids).[13–15] One way to explore the importance of such a process is to block the H dynamics in a complex as in the case of protonated tryptophan for which the excited lifetime is considerably longer when the H loss mechanism is blocked by a water molecule.[3] Thus if a longer lifetime is observed in the homo-dimers of DNA bases, it can be the sign of the importance of the H atom dynamics and this is one of the goal of the paper.

The characterization of the most stable protonated conformer/tautomer is usually obtained through combination of infrared spectroscopy and ab-initio calculations.[16–25] We have shown that electronic spectroscopy can be also used as a fingerprint to differentiate isomers and assign the most stable structures in the assumption that the thermal equilibrium is obtained in an ion trap, and that the oscillator strength, the Franck Condon factors and the fragmentation yield are known.[9,12,26]

In addition to the canonical Watson-Crick pairs, DNA base pairs can form non-canonical pairs[27] for which special hydrogen bonds are found, such as Cytosine$\cdots$H$^+\cdots$Cytosine ($C_2H^+$)[28] formed in DNA at acidic pH values. In this $C_2H^+$ base pair, two Cytosine residues of two parallel strands share a single proton on the nitrogen atoms in the 3-position. The study of isolated protonated base pairs will allow determining their absorption spectra that may be used to identify the non-canonical structures.

We present new experimental results on the protonated homo-dimers of pyrimidine DNA/RNA bases (cytosine, thymine and uracil) in gas phase. It should be mentioned that these spectra have never been recorded before. The goal of this work is to address the following questions:

i. Are the dimer electronic absorptions drastically different from the monomer absorptions as in the case of protonated benzene dimer compared to protonated benzene?[11] i.e. can the electronic spectroscopy be used to monitor the formation of protonated dimers?

ii. Is there a strong change in the excited state lifetime between the protonated monomers and dimers (monitored through spectral lifetime broadening)?

iii. Can the electronic spectra of the homo-dimers be assigned to specific tautomers?

## 2. Methods

### 2.a. Experimental

The setup has been described previously.[9,15] The electronic spectra of the homo-dimers of the protonated DNA/RNA bases were obtained via parent ion photo-fragmentation spectroscopy in a cryogenically-cooled quadrupole ion trap (Paul Trap from Jordan TOF Products, Inc.). The setup is similar to the one developed in several groups based on the original design by Wang and Wang.[29,30] The protonated ions are produced in an electrospray ionization source built at Aarhus University.[31] At the exit of the capillary, ions are trapped in an octopole trap for 90 ms. They are extracted by applying a negative pulse of c.a. 50 V and are further accelerated to 190 V by a second pulsed voltage just after the exit electrode. This time sequence of pulsed voltages produces ion packets with duration between 500 ns and 1 µs. The ions are driven by a couple of electrostatic lenses toward the Paul trap biased at 190 V so that the ions enter the trap gently avoiding fragmentation induced by collisions. A mass gate placed at the entrance of the trap allows selecting the parent ion. The Paul trap is mounted on the cold head of a cryostat (Coolpak Oerlikon) connected to a water-cooled He compressor. Helium as buffer gas is injected in the trap using a pulsed valve (General Valve) triggered 1 ms before the ions enter the trap as previously reported by Kamrath et al.[32] The ions are trapped and thermalized at a temperature between 20 and 50 K through collisions with the cold buffer gas. The ions are kept in the trap for several tens of ms before the photodissociation laser is triggered. This delay is necessary to ensure thermalization of ions and efficient pumping of the He buffer gas from the trap to avoid collision induced dissociation of the ions during the extraction towards the 1.5 m long time-of-flight mass spectrometer. After laser excitation, the ions are stored in the trap for a delay that can be varied between 20 and 90 ms before extraction to the TOF mass spectrometer. Full mass spectrum is recorded on a micro channel plates (MCP) detector with a digitizing storage oscilloscope interfaced to a PC. The photofragmentation yield detected on each fragment is normalized to the parent ion signal and the laser power. We can also detect the neutral fragments produced by photodissociation of the parent ions in the TOF, which allows to perform UV-UV hole burning experiment.[26]

The photo-dissociation laser is an OPO laser from EKSPLA, which has a 10 Hz repetition rate, 10 ns pulse width, a resolution of 10 $cm^{-1}$ and a scanning step of 0.02 nm. The laser is shaped to a 1 $mm^2$ spot to fit the entrance hole of the trap and the laser power is around 20 mW in the UV spectral region.

### 2.b. Calculations

*Ab initio* calculations have been performed with the TURBOMOLE program package,[33] making use of the resolution-of-the-identity (RI) approximation for the evaluation of the electron-repulsion integrals.[34] The equilibrium geometry of the protonated species in their ground states ($S_0$) have been determined at the MP2 (Møller-Plesset second order perturbation theory) level. Vertical excitation energies of the lowest excited singlet states have been determined at the RI-ADC(2)(second order Algebraic Diagrammatic Construction) level.[35] Due to numerous conical intersections between the nπ* and ππ* states,[36,37] the excited state optimization is not straightforward and failed in many systems as shown already for protonated adenine.[38] Calculations were performed with the correlation-consistent polarized valence double-zeta (cc-pVDZ) basis set.

## 3. Experimental Results

All the dimers show a very broad absorption band in the UV starting around 33500 - 35000 $cm^{-1}$ and extending over about 9000 $cm^{-1}$ (upper right parts of Figures 1, 2 and 3). For all of them, the protonated monomer is the only photofragment observed, which corresponds to breaking the weakest bond. Compared to the protonated monomers, the relative fragmentation yield is large and can reach 50%.

### *a) Protonated cytosine dimer*

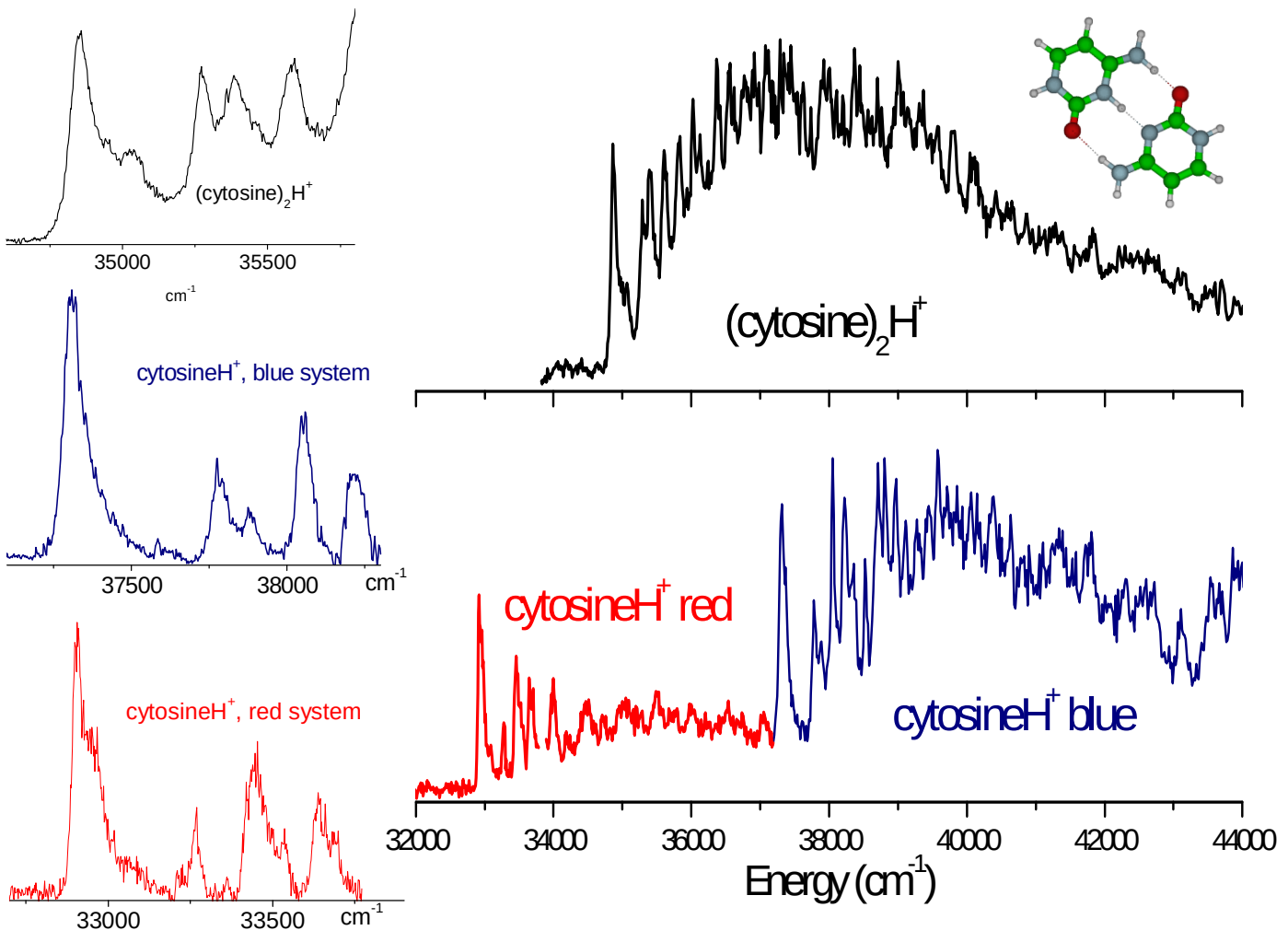

***Figure 1:*** *photo-fragmentation spectrum of the protonated cytosine dimer (cytosine)$_2$H$^+$ compared to protonated cytosine monomer cytosineH$^+$. Upper right: protonated cytosine dimer photo-fragmentation spectrum; lower right: protonated cytosine monomer photo-fragmentation spectrum. The red band system is assigned to the K$^+$ tautomer and the blue band system to the E$^+$ tautomer. Left: enlarged view of the band origins; upper left: (cytosine)$_2$H$^+$, middle left: E$^+$ monomer (blue system), lower left: K$^+$ monomer (red system). All the bands are intrinsically broad*

For the protonated cytosine monomer, two band systems have been observed, the red one (32900 $cm^{-1}$) being assigned to the ππ* ← ππ transition of the Keto ($K^+$) tautomer and the blue band (37305 $cm^{-1}$) to the transition of the Enol ($E^+$) tautomer.[12] The spectrum of the protonated dimer of cytosine (Figure 1) consists of well defined vibronic bands starting with a 0-0 transition at 34850$cm^{-1}$, between the origins of the two bands systems of protonated cytosine monomers. The vibrational progression is similar to the one observed in the protonated cytosine spectra, which indicates that the intramolecular vibrations are not heavily perturbed in the proton bond dimer. The bands are broad, which can be due to a short excited state lifetime or spectral congestion.

### *b) Protonated thymine dimer*

The protonated thymine dimer exhibits a spectrum extending over more than 9000 cm$^{-1}$ (Figure 2 upper right) with sharp vibrational transitions clearly observed at the origin of the transition (33625 cm$^{-1}$, 4.17 eV) (upper left part of Figure 2). The origin of this progression is red-shifted by -3775 cm$^{-1}$ from the origin of the transition of the protonated thymine monomer (Enol-Enol (EE$^+$) tautomer at 37400 cm$^{-1}$, 4.63 eV).[12] The Enol-Keto (EK$^+$) tautomer of the protonated thymine monomer has not been observed in our previous experiment but the origin of its electronic transition has been calculated to be around 28600 cm$^{-1}$.[12] Thus, the electronic absorption of the protonated dimer lies between the transitions of the two tautomers of protonated thymine. In protonated thymine, the 0-0 transition is quite broad, which has been assigned to a very short excited state lifetime,[12] whereas in the protonated dimer, the vibrational structure near the origin is well resolved with vibrational bandwidths of 10 cm$^{-1}$, which is the spectral resolution of the laser. The contrast is quite astonishing, if one compares the spectrum of the protonated monomer in which the bands are quite broad (100 cm$^{-1}$) to the spectrum of the protonated homo-dimer in which the bands are at least ten times narrower (left side of Figure 2) near the origin. There is no reason to have a simpler vibrational structure in the complex than in the monomer, so the proposition that the width of the bands in protonated thymine is due to lifetime broadening seems to be strengthened. In other words, the complexation seems to inhibit the non-radiative decay.

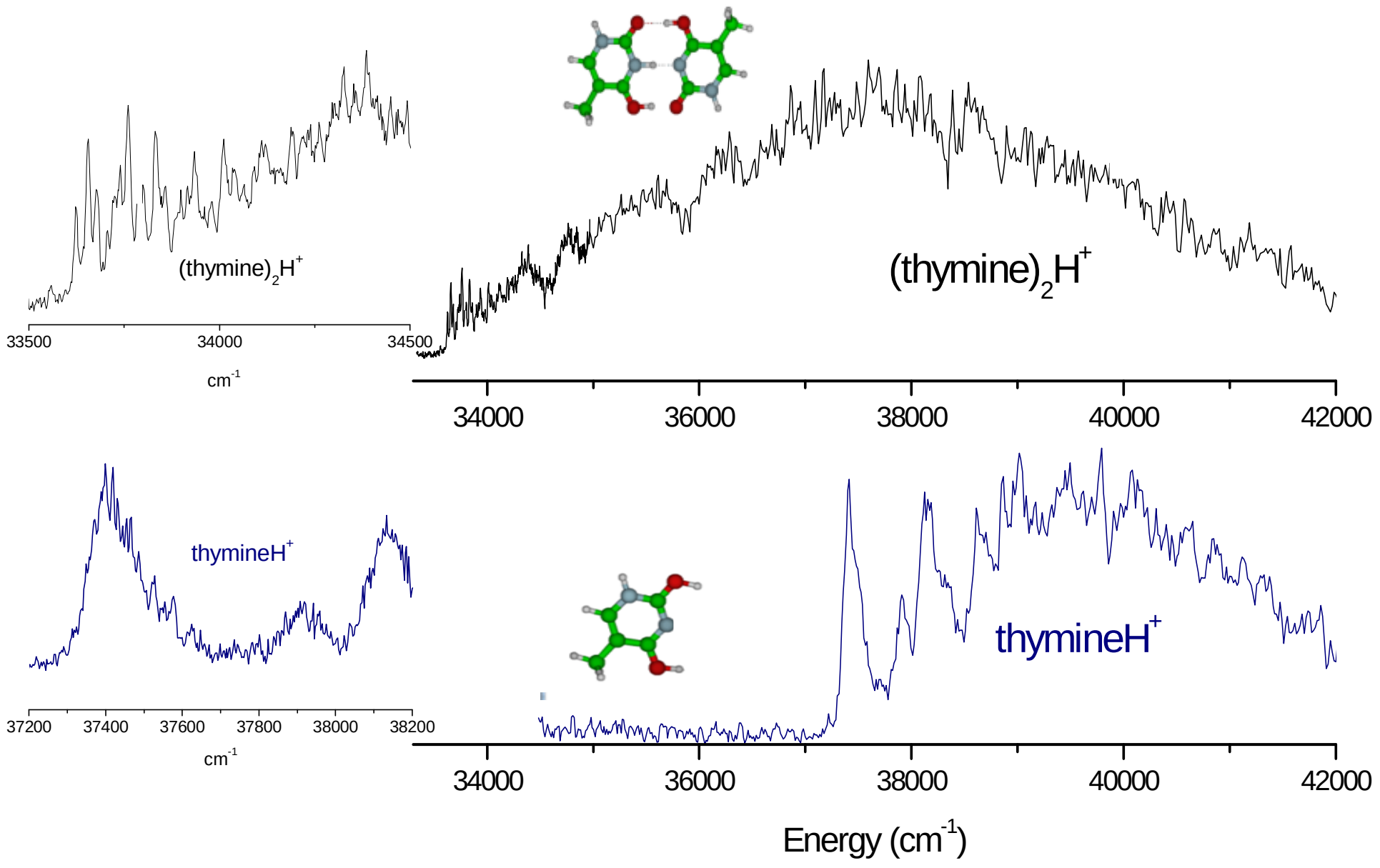

***Figure 2:*** *Comparison between the photo-fragmentation spectra of protonated thymine dimer (thymine)$_2$H$^+$ and protonated thymine monomer thymineH$^+$. Upper right: protonated thymine dimer photo-fragmentation spectrum; lower left: protonated thymine photo-fragmentation spectrum (only the EE$^+$ structure has been observed); left: enlarged view of the band origins: Upper trace: for the protonated thymine dimer, the vibrational bandwidths are laser limited, lower trace: at the same scale, the bands of the protonated monomer are considerably broader.*

The vibrational progression is not easy to assign due to the lack of resolution of the laser and the small vibrational frequency spacing (around 30 cm$^{-1}$). From calculation of the ground state frequencies of the protonated thymine dimer, these vibrational modes can be assigned to intermolecular out-of-plane vibrations, the in-plane intermolecular frequencies being in the order of 120 $cm^{-1}$. These modes being active in the spectrum, the excited state should have a slightly non planar structure.

### *c) Protonated uracil dimer*

In protonated uracil as in protonated cytosine (monomers), two electronic band systems have been observed,[12] which have been assigned to transitions of two tautomers, the $EE^+$ and $KE^+$, through calculations[12] and UV-UV hole burning experiments.[26] The red isomer assigned to the $KE^+$ tautomer shows sharp vibrational bands whereas for the blue one ($EE^+$ tautomer) the bands are much broader, which has been assigned to lifetime broadening.

For the protonated uracil dimer (Figure 3), in contrast with the thymine case, the band contour in the vicinity of the 0-0 transition is very broad, as for the $EE^+$ protonated monomer. The bandwidth can be due to spectral congestion or to lifetime broadening but it is not clear why there should be more spectral congestion in this system than in the protonated thymine dimer.

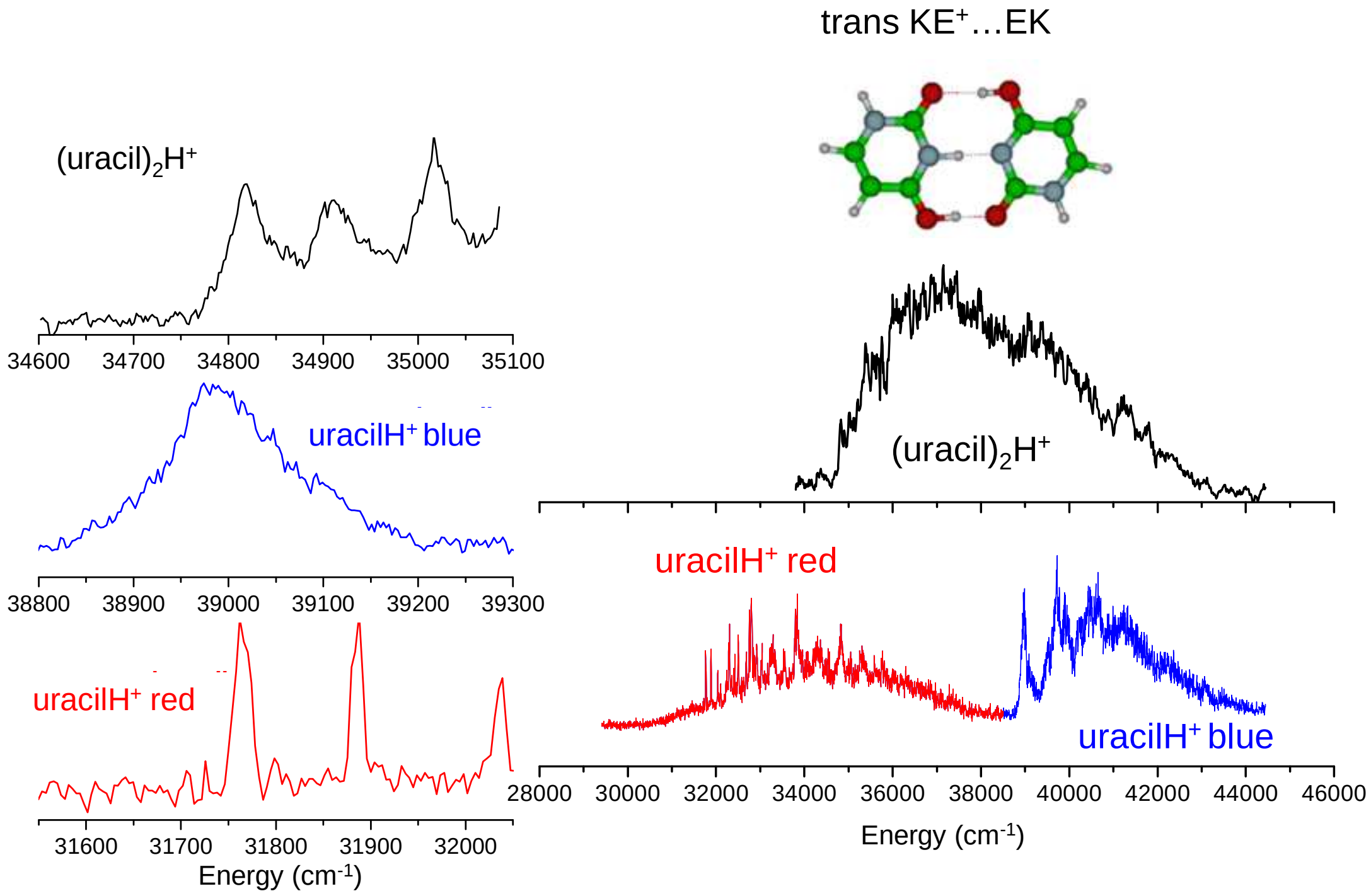

***Figure 3:*** *Comparison between the photo-fragmentation spectra of protonated uracil dimer and protonated uracil monomer. Upper right: protonated uracil dimer photo-fragmentation spectrum. Lower right: protonated uracil photo-fragmentation spectrum. The red part is assigned to the* $KE^+$ *tautomer and the blue part to the* $EE^+$ *tautomer. Left enlarged view of the origin region: upper panel: protonated uracil dimer, middle panel:* $EE^+$ *monomer, lower panel:* $KE^+$ *monomer. The bands of the protonated dimer are broader than the bands of the* $KE^+$ *uracil tautomer (red band system).*

## 4. Calculations and discussion

The most strongly bond dimers are those in which the number of hydrogen bonds is maximized. The ground states of π-bonded conformers have been calculated to be at least 1 eV higher in energy than that of σ-bonded conformers. For the protonated cytosine dimer, only one isomer allows three hydrogen bonds but for thymine and uracil a few isomeric structures can be obtained. In Table 1 are summarized the ground state energies, the vertical excited states energy ($S_1$, $S_2$ and the charge transfer (CT) state), the difference between the $S_1$ vertical values and the experimental measurement and the adiabatic $S_1$ transition energies calculated in Cs symmetry.

**Table 1:** Calculations of the ground and excited states for the protonated pyrimidine dimers. All the energies are in eV.

| Dimer | Name | $S_0$ [(a)] | $S_0$-$S_1$ exp | $S_0$-$S_1$ cal [(b)] | $S_0$-$S_2$ [(b)] | exp-cal [(c)] | CT [(d)] | opt a' [(e)] |
|---|---|---|---|---|---|---|---|---|
| **Cytosine** 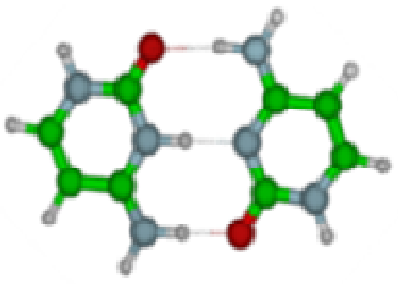 | Cis $K^+$...K [(f)] | 0 | 4.32 | 4.91 | 4.96 | -0.59 | 5.73 | 4.44 |
| **Uracil** 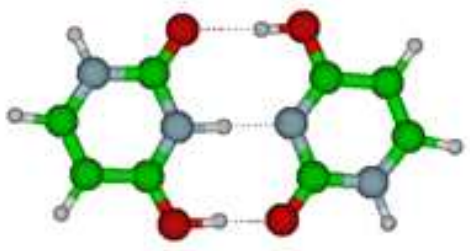 | Trans [(g)] $KE^+$...EK | 0 | 4.32 | 4.83 | 4.93 | -0.51 | <5.8 | 4.35 |
| 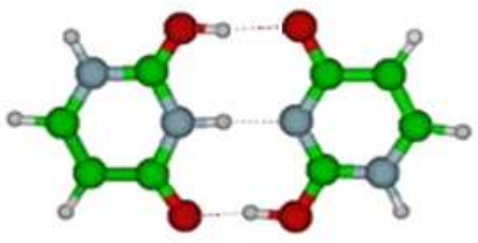 | Trans $EK^+$...KE | PT [(h)] | | | | | | |
| 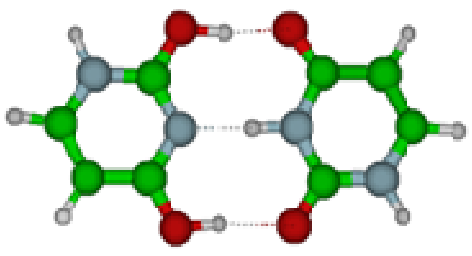 | Trans $EE^+$...KK | 0.07 | | 5.29 | 5.33 | -0.97 | 6.12 | PT [(i)] |
| 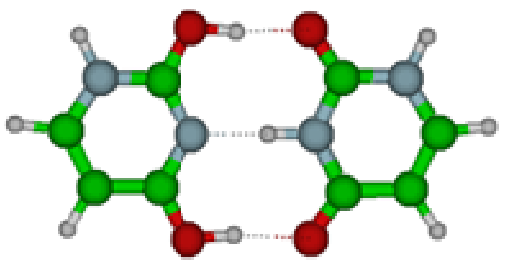 | Cis [(g)] $EE^+$...KK | 0.084 | | 5.44 | 5.47 | -1.12 | | |
| 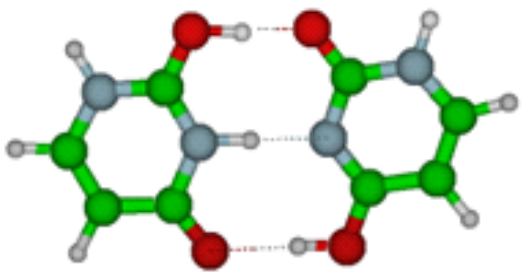 | Cis $EK^+$...KE | 0.124 | | 4.94 | 4.99 | -0.62 | | |
| 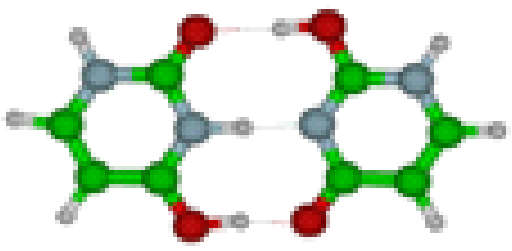 | Cis $KE^+$...EK | 0.15 | | 4.90 | 5.20 | -0.58 | | |
| **Thymine** 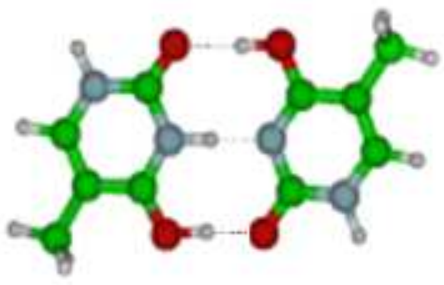 | Trans $KE^+$...EK | 0 | 4.17 | 4.68 | 4.8 | -0.51 | 5.85 | 4.23 |
| 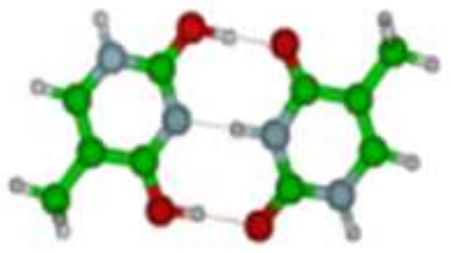 | Trans $EE^{+}$...KK | 0.031 | | 5.17 | 5.19 | -1.00 | 5.77 | PT[(i)] |
| 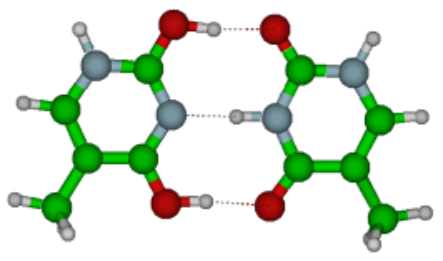 | Cis $EE^{+}$...KK | 0.04 | | 5.19 | 5.39 | -1.02 | | |
| 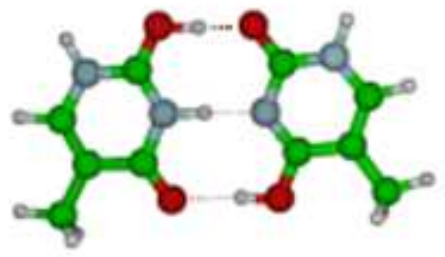 | Cis $EK^{+}$...KE | 0.084 | | 4.84 | 4.99 | -0.67 | | |
| 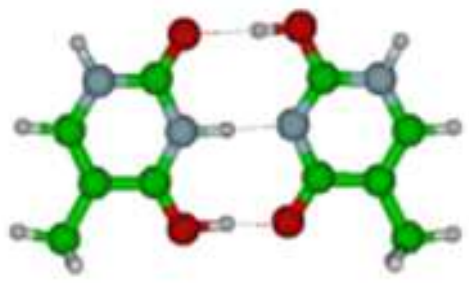 | Cis $KE^{+}$...EK | 0.108 | | 4.73 | 5.25 | -0.56 | | |
| 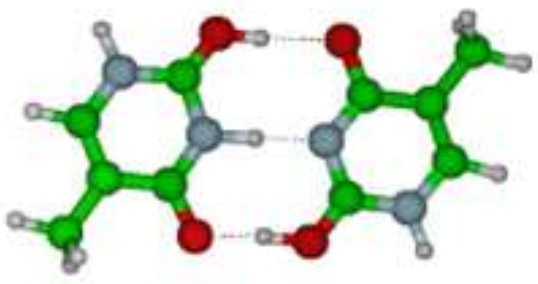 | Trans $EK^{+}$...KE | 0.19 | | 5.41 | 5.49 | -1.24 | | |

*(a) the ground state energies are calculated at the MP2/cc-pVDZ level*
*(b) the excited states are calculated at the ADC(2)/cc-pVDZ level*
*(c) difference between the experimental band origin (exp) and the $S_1$ calculated vertical energy (cal)*
*(d) CT means charge transfer state*
*(e) opt a' means adiabatic $S_1$ transition energies calculated in Cs symmetry*
*(f) the structure is the same as calculated in references*[39,40]
*(g) trans and cis isomers refers to the relative orientation of the nitrogen atoms, where the glycosidic bond takes place in nucleosides, of one moiety with respect to the other moiety.*
*(h) PT means that the initial trans $EK^{+}$…KE structure converges in the ground state to the trans $EE^{+}$...KK isomer, i.e. there is a proton transfer*
*(i) PT means that the initial $EE^{+}$…KK structure converges in the excited state to the lowest excited state of the $EK^{+}$...EK isomer, i.e. there is a proton transfer.*

### a) Ground state structure

One expects that the homo-dimer is built in bringing together the most stable tautomeric structures of both the neutral and the protonated monomers. For example, in the case of the protonated uracil monomer, the most stable structure is the $EE^{+}$ isomer, which lies 0.08 eV below the $KE^{+}$ isomer,[12,16,41] while for neutral uracil, the Keto-Keto (KK) form is the most stable one, lying 0.5 eV lower in energy than the Enol-Keto (EK) tautomer. When the two species are separated, the protonated $EE^{+}$ plus neutral KK system has an energy 0.6 eV lower than the protonated $KE^{+}$ plus neutral EK system (Figure 4). But the $KE^{+}$….EK dimer is calculated to be the most stable structure (Figure 4) and has a calculated electronic transition that corresponds to the experimental observation (Table 1). The same observation is also valid for the protonated thymine dimer. The stronger stability of the $KE^{+}$….EK structure can be seen as a nice example of the H bond resonance effect.[42,43] This shows also that the most stable structure of the dimer cannot be simply deduced from the monomer stability as shown also in neutral systems.[44,45]

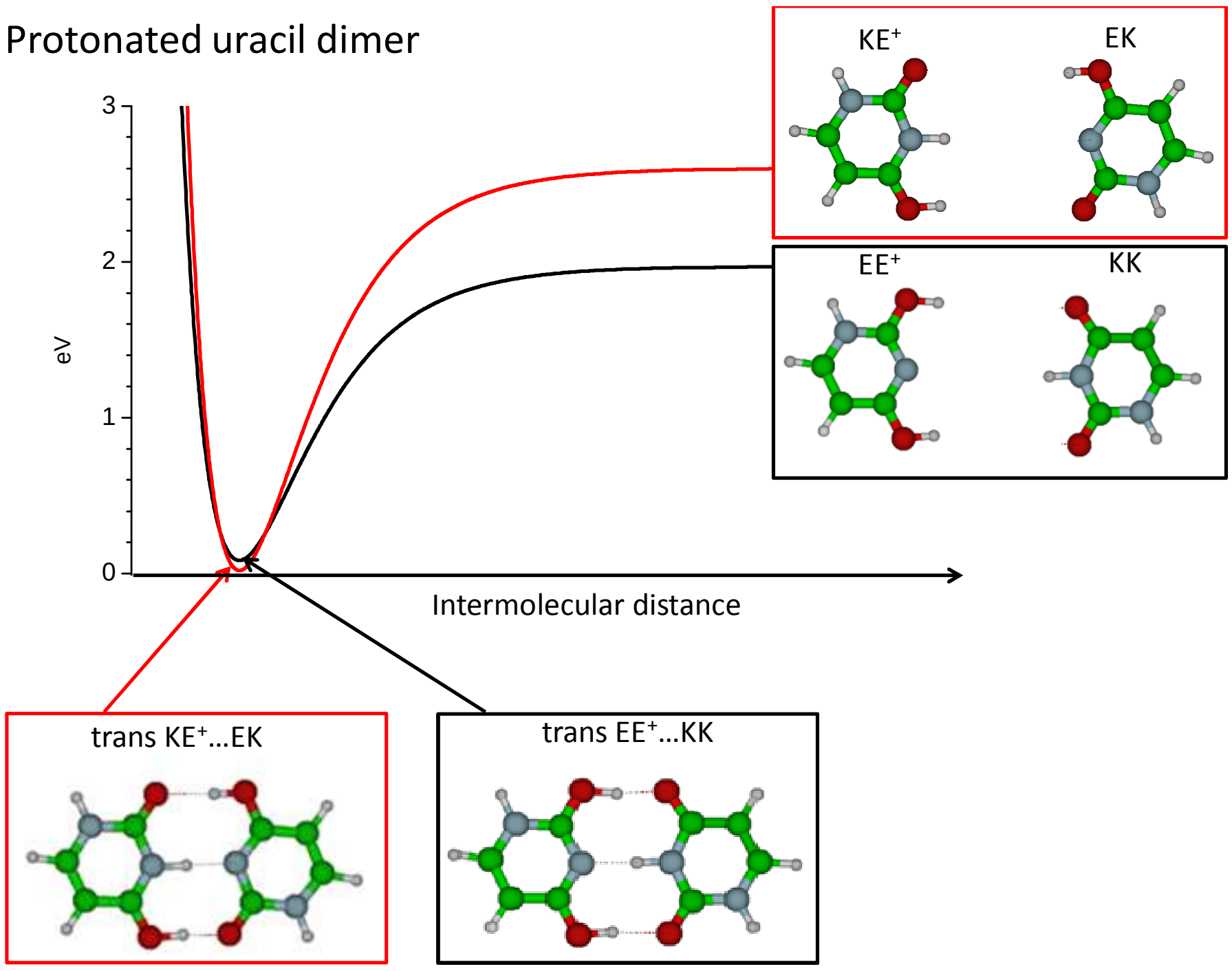

***Figure 4:*** *schematic of the dissociation potential curve for the protonated uracil dimer. At long distance the EE$^+$…KK structure is by far the most stable one but at the equilibrium geometry the KE$^+$…EK is the most stable.*

For uracil, in the ground state of the trans KE$^+$….EK structure, calculations at the MP2 level localize the proton on one side (NH distance=1.07 Å). The transition state for which the H atom is at equal distance between the two Nitrogen (NH=1.36 Å in $C_{2v}$ symmetry) is higher in energy by 2150 cm$^{-1}$, which is slightly more than the zero point energy of the NH vibration. Thus the proton is probably tunneling from one position to the other, i.e. the protonated dimer in the ground state has a Zundel type structure.

The KE$^+$….EK and EE$^+$…KK structures can also interconvert by proton transfer from one oxygen on the protonated KE$^+$ moiety that becomes the KK tautomer, to one oxygen atom of the neutral EK part that becomes protonated EE$^+$ isomer. The barrier for this PT in the ground state is calculated between 700 cm$^{-1}$ (minimum energy path at the DFT level) and 4000 cm$^{-1}$ (non relaxed O-H distance scan at the MP2 level ) (Figure 5). The barrier is probably not high enough to localize the proton on one isomer or the other. Thus, the notion of specific isomer has no real meaning, as already discussed by other authors.[46]

The case of protonated thymine and cytosine dimers should be quite similar, only small variations of the barrier heights are expected.

### *b) Assignment of the electronic transitions*

As for the protonated monomers, the assignment of the isomers can be obtained from the ground state energy and the excited state vertical transition.[12] Due to the low temperature of the trap, only the lowest energy isomers (less than 0.1 eV above the most stable one) are considered for excited state calculations. In many aromatic molecules, the excited state vertical energies are consistently calculated ~ 0.5 eV higher in energy than the experimental ones, this effect being due to the variation of the energy upon the excited state optimization.[7,8,10,12,47] This is what we also found in this work for the lowest energy isomers of the homo-dimers (Table 1). In the protonated uracil dimer, the $EE^+$…KK isomer, which is only 0.07 eV higher in energy than the $KE^+$….EK isomer in the ground state, cannot be responsible for the absorption at 4.32 eV since it has a vertical excitation energy at 5.3 eV, which would correspond to an adiabatic $S_1 \leftarrow S_0$ transition at 4.8 ± 0.2 eV, rather far from the observed experimental band origin. This isomer may be present but hidden in the broad continuous absorption at higher energy.

In most of the monomers of DNA bases (neutral or protonated), the excited state optimizations are not possible due to the out-of-plane deformation of the ring, which leads to many conical intersections where the calculation fails.[36,38] For the protonated dimers, the ground state structures are planar and for the most stable structures of Table 1, excited state optimizations have been performed with a Cs geometry constraint. The Cs adiabatic energies (last column in Table 1) are very close to the experimental values indicating that the excited states are indeed planar or just slightly out-of-plane distorted.

### *c) Electronic transition for the different tautomers of the homo-dimers*

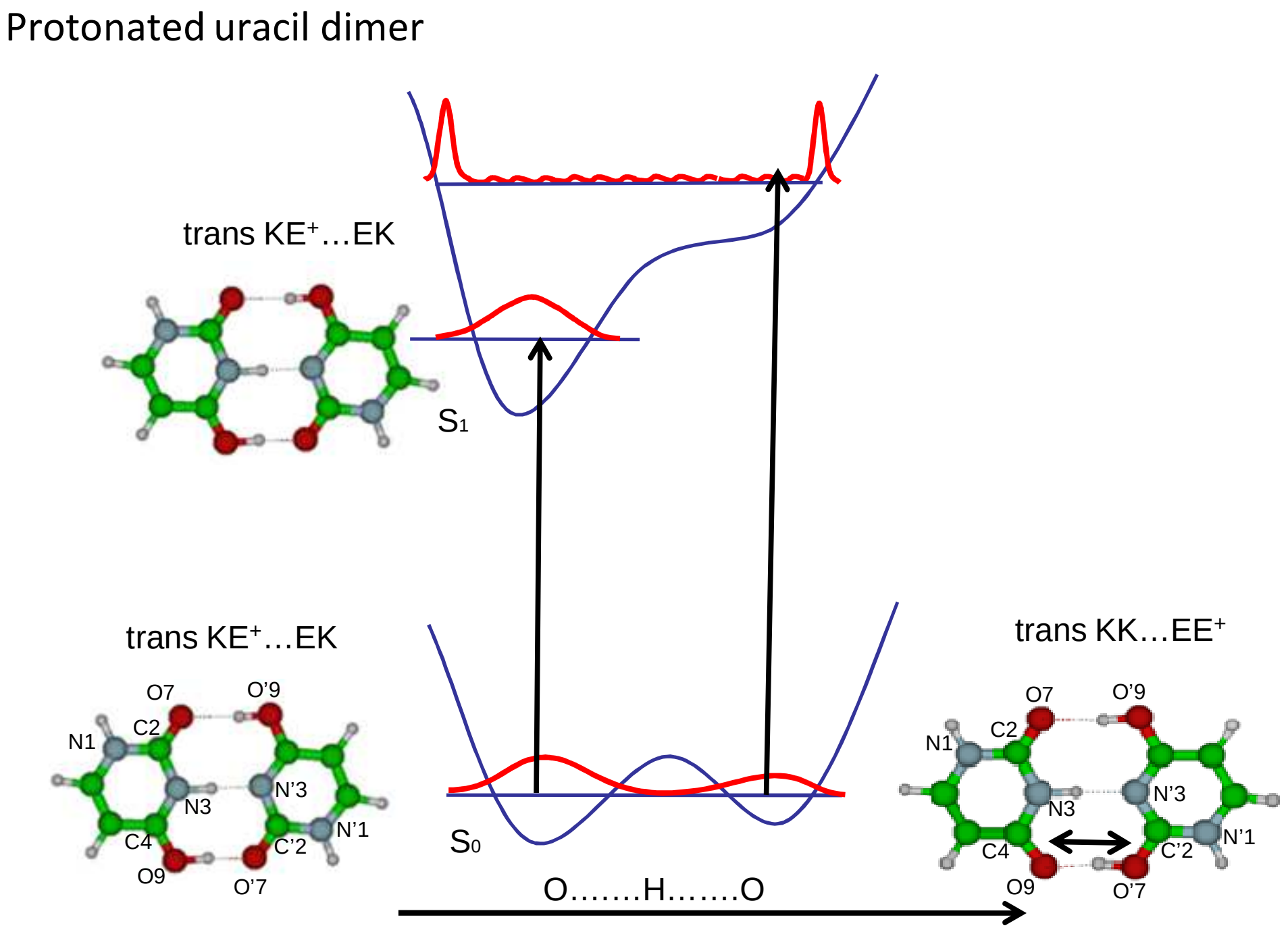

***Figure 5:*** *scheme of the potential energy function along the OH..O coordinate in the protonated uracil dimer. The blue horizontal lines indicate v=0 level, which is near the barrier in the ground state, indicating that both tautomers cannot be discriminated. In the excited state one tautomer has a quite significantly lower energy than the other. Excitation at the origin of the* $S_1 \leftarrow S_0$ *transition localizes the proton in only one tautomeric form.*

For the protonated monomers, both $E^+$ and $K^+$ tautomers of cytosine or $KE^+$ and $EE^+$ tautomers of uracil have been experimentally observed or calculated for thymine.[12] The protonated $K^+$ or $KE^+$ forms, when observed, have an excited state band origin red shifted by 0.5 to 1 eV as compared to the $EE^+$ forms. The Cs excited state optimization of the first $\pi\pi^*$ (A' in Cs geometry) of the $EE^+$….KK isomer leads to the $KE^+$….EK isomeric structure without barrier on the PT reaction path (Table 1). The potential curves along the O-H…O coordinate are schematically presented in Figure 5. The difference in the vertical transition energies between both species is around 1 eV, thus, the eventual barrier vanishes in the excited state. Excitation of the system starting from the $EE^{+}$…KK structure will reach a high energy part of the potential where the density of states is high and the FC factors will be weak. Thus this absorption will be buried in the continuum observed at high energy and only the isomer with the lowest electronic transition can be detected clearly.

It should be noted that although in the ground state the vibrational wavefunction is not localized specifically on one of the tautomers, the excitation at the origin of the $S_1 \leftarrow S_0$ transition localizes the proton in only one tautomeric form since the excitation energies and Franck Condon factors are quite different for both isomers. Therefore, although the proton is delocalized in the ground state and the notion of specific isomer is meaningless in this state, it is justified in the excited state and from a theoretical point of view, it is the only way to get electronic properties.

### *d) Nature of the excited state*

For most of the dimers (in particular the most stable ones), the first excited state is a transition localized on the protonated entity and corresponding to a ππ* excitation. The second excited state corresponds to an electronic transition on the neutral species. In a previous study on the electronic absorption of protonated benzene dimer, the first electronic transition was due to a charge transfer transition from the HOMO on the neutral part to the LUMO on the protonated ion.[11] This was giving rise to an absorption strongly shifted to the low energy region i.e. in the visible. A first electronic transition with a strong CT character leading to a red shifted absorption was also observed for protonated polycyclic aromatic molecules.[10] In the protonated pyrimidine base dimers, the CT transitions (nπ* and πσ* states) are at the best, the third electronic transition, 1 eV higher than the locally excited ππ* transitions. In the protonated benzene dimer, the charge is clearly localized on the protonated species (100%) whereas in the protonated pyrimidine base homo-dimers the charge is delocalized at 20% on the neutral part of the systems in the ground state.

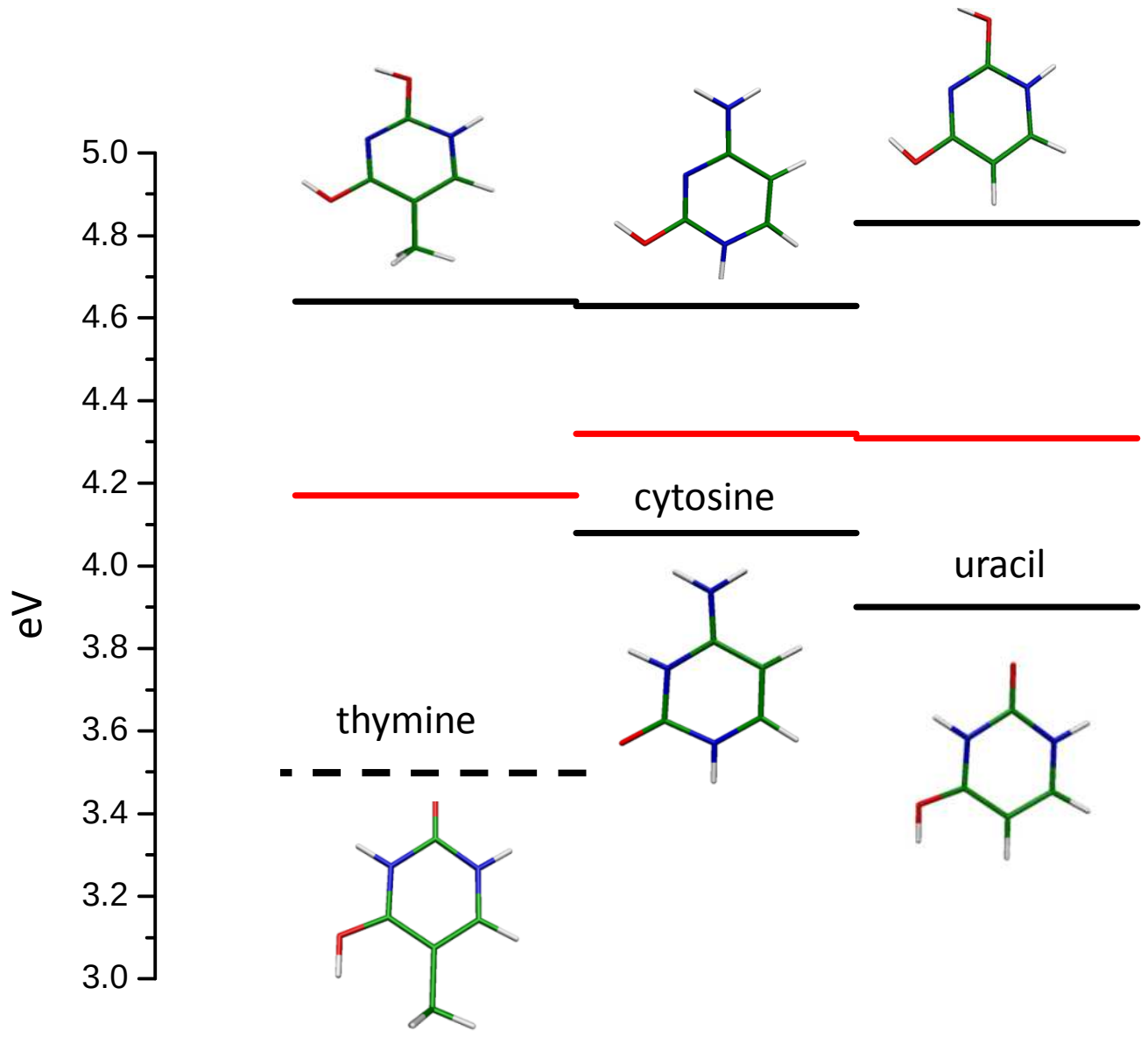

***Figure 6 :*** *Experimental energies of the electronic transitions for the different tautomeric structures of the protonated monomers (in black) and electronic transitions of the protonated homo-dimers (in red). For protonated thymine the value of the EK$^{+}$ transition (dotted line) is not observed but calculated.*

The 0-0 transition energies of the protonated homo-dimers are in between the transition energies of the E$^{+}$ (or EE$^{+}$) and K$^{+}$ (or EK$^{+}$) tautomers of the protonated monomers as seen in Figure 6, which could be due to the delocalization of the proton between the two moieties of the dimers (Figs. 4 and 5).

### *e) Line broadening and excited state lifetime*

The variation of the vibrational bandwidths has already been observed in the protonated DNA/RNA bases.[12] As an example in the protonated uracil monomer, the KE$^{+}$ structure presents narrow vibrational bands (12 cm$^{-1}$ FWHM) whereas the EE$^{+}$ structure exhibits very broad bands (100 cm$^{-1}$ FWHM).

In the protonated thymine dimer, a vibrational progression in a low frequency mode (33 cm$^{-1}$) is clearly observed, whereas in the protonated monomer, the bands have a FWHM of (91$\pm$20) cm$^{-1}$. The broadening in the protonated monomer can be due to a short excited state lifetime or to spectral congestion with low frequency mode separated by less than the experimental resolution i.e. 10 cm$^{-1}$. This second possibility is very unlikely since such low frequency modes are only expected for very floppy molecules and should be also present in the dimer. Thus it is reasonable to assume that the broadening in the protonated monomer is due to the excited state lifetime, which implies that the excited state lifetime of the protonated thymine dimer is longer than that of the monomer. The reason for a shorter lifetime in the protonated monomer than in the dimer is not clear.

One possibility to explain the lifetime difference is that we are not comparing the same conformers in the monomer and in the dimer. In the protonated dimer, the protonated part is of the EK$^{+}$ form whereas the conformer observed for the protonated monomer is the EE$^{+}$ form. Unfortunately, the EK$^{+}$ protonated monomer has not been observed in the trap and direct comparison cannot be done. This would imply that the $EK^+$ monomer has a long lifetime as in the protonated uracil monomer.

Another possibility could be that in the monomer the short lifetime is controlled by the H loss mechanism mediated by a πσ* NH/OH dissociative state as in many protonated aromatic molecules (tryptophan, aromatic amines…)[15,48] or neutral systems (phenol, indole…).[49] In these systems the complexation with a molecule preventing the direct H loss increases considerably the excited lifetime.[3,49] However, the signature of this mechanism in the protonated monomer would be the H loss channel, which has not been observed experimentally. One explanation could be the coupling of the πσ* state with the ground state, leading to internal conversion.[15]

For the other two dimers, the vibronic bands are broad, either the lifetime is short or the spectral congestion is more severe than in the protonated thymine dimer. Laser spectroscopy with higher resolution or pump-probe fs experiments are needed to solve this issue. A comparison with neutral systems would be interesting but such data for pyrimidine bases seems to be still missing.[50,51]

If we assume that the bandwidths in the protonated uracil dimer spectrum are due to short lifetimes, one could wonder why the lifetimes of thymine and uracil dimers are so different? We searched for some explanation in the nature of the orbitals, and in the energy of the nπ* as compared to the ππ* but we found no obvious explanation. Maybe the difference of lifetime is not as large as it seems: the lifetime for the protonated thymine dimer is at least 1 ps (this is the lower limit to observe resolved spectra) and for the protonated uracil dimer the lifetime is of the order of 200 fs. This decrease of lifetime by a factor of ~5 from the protonated thymine dimer to the protonated uracil dimer would be enough to wipe out the vibrational structure due to very low frequency. A very small change in the potential energy surface (position of the conical intersection, coupling elements) would be sufficient to explain the observed experimental behavior.

## Conclusions

We have presented the electronic photo-fragmentation spectra of the protonated pyrimidines DNA/RNA homo-dimers. The absorption of the first exited state is relatively structured indicating that the lifetime of the excited states is longer than a few hundreds of femtoseconds, and even longer for the protonated thymine dimer. The case of the protonated thymine dimer is surprising since sharp vibrational bands are observed. The electronic absorptions of the dimers lie between the absorptions of the different tautomers of the protonated monomer ($E^+/EE^+$ and $K^+/KE^+$) and calculations show that the transition corresponds to a localized excitation on the protonated part. The electronic shift of the bands of the dimers with respect to that of the monomers is larger than 0.2 eV which means that the electronic absorption might be used for monitoring the formation of dimers in condensed medium.

The calculations show that several tautomers of dimers exist in the ground state, probably due to the hydrogen bond resonance effect.

## Acknowledgements

This works was supported by ECOS-MinCyT cooperation program (A11E02) the ANR Research Grant (ANR2010BLANC040501), FONCyT, CONICET and SeCyT-UNC. We acknowledge the use of the computing facility cluster GMPCS of the LUMAT federation (FR LUMAT 2764).

## References

† In all the manuscript, we use the notation $E^+$ or $K^+$ for the protonated cytosine monomer to indicate that the proton is on the oxygen or on the nitrogen (N3), respectively. For protonated thymine and uracil monomers, we use the notation $EE^+$ to indicate that the hydrogen and proton are on the two oxygen atoms and $EK^+$ when the hydrogen and proton are on one oxygen and on nitrogen (N3) atoms.

We represent the dimers with the protonated moiety on the left. We use the notation $EE^+$…KK or $KE^+$…EK for the protonated thymine and uracil homo-dimers to indicate the hydrogen and proton positions on the protonated moiety…the hydrogen position on the neutral moiety.

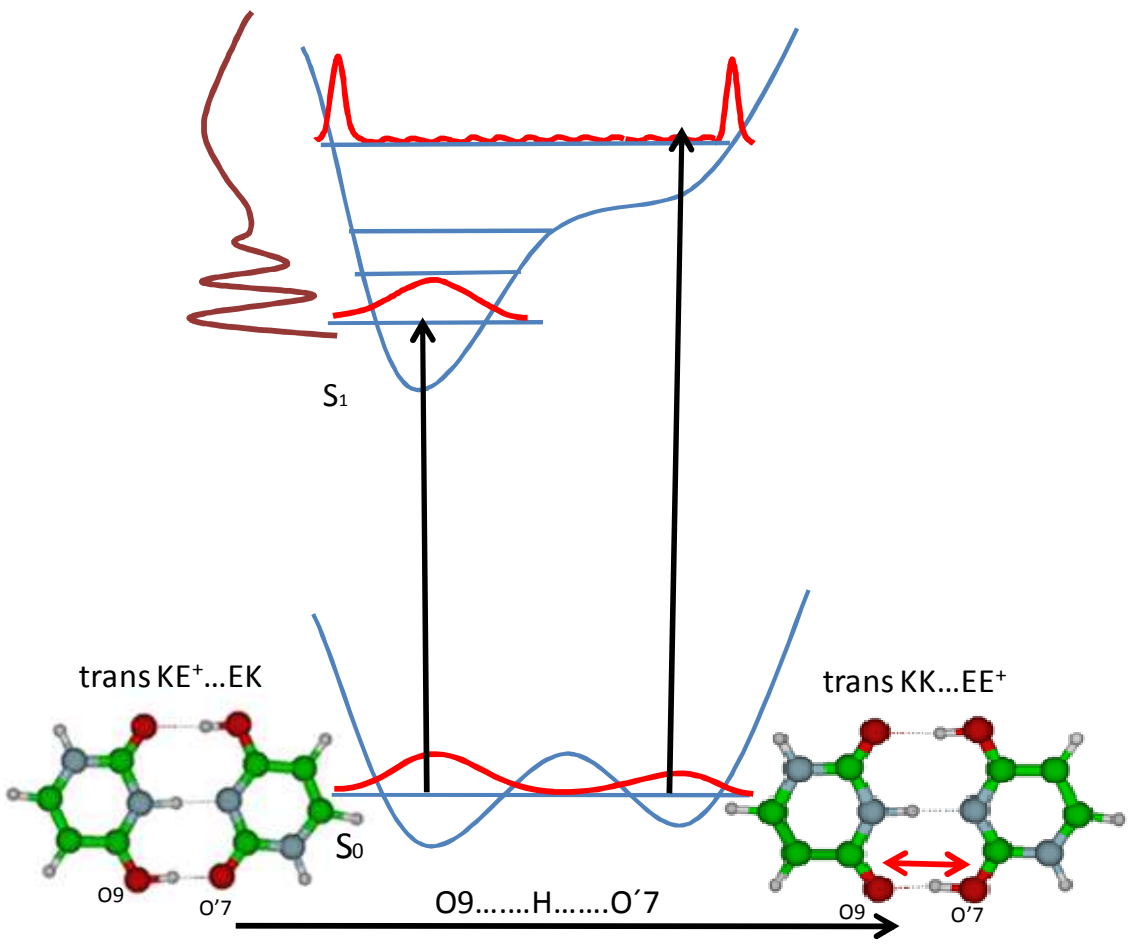